\documentclass[twocolumn,secnumarabic,amssymb, amsmath, nofootinbib,tightenlines,
nobibnotes, aps, prl,epsfig]{revtex4}
\usepackage{graphicx}
\usepackage{dcolumn}
\usepackage{bm}
\begin{document}
\preprint{APS/123-QED}
\title{Physical limits in the Color Dipole Model Bounds}

\author{G.R.Boroun}%
 \email{grboroun@gmail.com; boroun@razi.ac.ir }
\affiliation{ Physics Department, Razi University, Kermanshah
67149, Iran}
\date{\today}
\begin{abstract}
The ratio of the cross sections for the transversely and
longitudinally virtual photon polarizations,
$\sigma^{\gamma^{*}p}_{L/T}$, at high photon-hadron energy
scattering is studied. I investigate the relationship between the
gluon distribution obtained using the color dipole model and
standard gluons obtained from the
Dokshitzer-Gribov-Lipatov-Altarelli-Parisi (DGLAP) evolution and
the Altarelli-Martinelli equations. It is shown that the color
dipole bounds are dependent on the gluon distribution behavior.
This behavior is considered by  the expansion and Laplace
transform methods. Numerical calculations and comparison with the
color dipole model (CDM) bounds can indicate the range of validity
of
this method at small dipole sizes, $r\sim 1/Q{\ll}1/Q_{s}$.\\
\end{abstract}
 \pacs{***}
\maketitle
\subsection{1. Introduction}
Dipole model provides a convenient description of deep inelastic
scattering (DIS) at small $x$. In this region, partons  in the
proton form a dense system. Processes such as mutual interaction
and recombination leads to the saturation of the total cross
section. We know that the DIS cross section is factorized  into a
light-cone wave function. As the virtual photon splits up into a
quark-antiquark pair (dipole). Usually, this contribution
($\gamma^{*}{\rightarrow}q\overline{q}$) is defined  as a
convolution of the infinite momentum frame wave function with the
pQCD calculable coefficient functions. These coefficient functions
describing the short distance propagation of the particles between
two virtual photon vertices. Many of experimental provided at HERA
have been analysis in terms of this model which the dipole picture
acts like a quark-antiquark dipole [1-4]. We need to know that the
lifetime of the fluctuation of the photon into the color dipole is
much larger than the typical timescale of the dipole-proton
interaction [5]. In this reference frame, the photon splits into
$q\overline{q}$-dipole and then interacts with proton as
$q\overline{q}$
lifetime is about $1/x$ times longer than time of interaction with proton.\\
The saturation region is approached when the reaction is mediated
by multi-gluon exchange. This aspect of saturation is closely
linked to unitarity. Indeed the growth of the gluon density is
slowed down at very small $x$ by gluon -gluon recombination ($g+g
\rightarrow g$). In this region the gluon density in hadrons can
become nonperturbatively large. It means that it is the regime of
gluon saturation. This is the origin of the shadowing correction
in pQCD interactions [1]. One of the implementation of multiple
scattering in colour dipole model  is based on the Glauber-Mueller
(GM) eikonal approach [2], which is used in
Golec-Biernat-W$\mathrm{\ddot{u}}$sthoff (GBW) model [3]. In this
approach the multiple colour dipole scatters are independent of
each other which describing the multi-gluon density in the proton
by the b-Sat model [4]. As the large saturation effects are
required to transition from the hard Pomeron behavior at small
dipole sizes to soft Pomeron behavior at large dipole sizes [6].\\
An effective field theory describing the small-$x$ regime of QCD
is the color glass condensate (CGC) [7,8,9]. This model is based
on the Balitsky-Kovchegov (BK) [10] non-linear evolution equation
and improves the Iancu-Itakura-Munier (IIM) dipole model [11]. The
non-linear effects appear in CGC formalism when the gluon density
becomes large. Indeed in high energy collisions which $x$
decreases, the number of gluons increase. When gluons are highly
coherent at infrared scale, the gluon saturation leads to the
Glasma [8,9,10,11]. Indeed the Glasma is matter produced from the
CGC  after a collision. After the collisions, Glasma is formed in
the region between the two sheets of colored glass. At high energy
scattering with evolution and recombination of gluon density, one
can probes the number of gluons with a given $x$ and transverse
momenta $k_{\bot}\leq Q$ as the gluon number is defined by
\begin{eqnarray}
x\frac{dN_{g}}{dx}(Q^{2})=\frac{\alpha_{s}C_{R}}{\pi}\int_{\Lambda^{2}_{QCD}}^{Q^{2}}
\frac{dk_{\bot}^{2}}{k_{\bot}^{2}}=\frac{\alpha_{s}C_{R}}{\pi}\ln(\frac{Q^{2}}{\Lambda^{2}_{QCD}})
\end{eqnarray}
where $\Lambda_{QCD}$ is the QCD cutoff and $C_{R}$ is the
$SU(N_{c})$ Casimir operator [9]. The HERA data collected on the
inclusive $\gamma^{*}p$ cross section for $x{\leq}0.01$ indicate a
scaling as a function  of the ratio $Q^{2}/Q^{2}_{s}(x)$. Which
$Q^{2}_{s}(x)=Q_{0}^{2}(x_{0}/x)^{\lambda}$ is the saturation
scale with dimensions given by a fixed reference scale $Q_{0}^{2}$
[12]. Indeed saturation scale is a border between dense and dilute
gluonic system. For $Q^{2}<Q_{s}^{2}$ the linear evolution is
strongly perturbed by nonlinear effects, while for
$Q^{2}>Q_{s}^{2}$ the linear evolution is dominated and evolution
of parton densities is governed by the DGLAP equations.\\
 The proton structure
function $F_{2}$ and the longitudinal structure function $F_{L}$
can be can be written in terms of  $\gamma^{*}p$ cross section as
follows
\begin{eqnarray}
F_{2}(x,Q^{2})&=&\frac{Q^{2}}{4\pi^{2}\alpha_{em}}[\sigma_{L}^{\gamma^{*}p}(x,Q^{2})+\sigma_{T}^{\gamma^{*}p}(x,Q^{2})]\\
F_{L}(x,Q^{2})&=&\frac{Q^{2}}{4\pi^{2}\alpha_{em}}\sigma_{L}^{\gamma^{*}p}(x,Q^{2})
\end{eqnarray}
where $\alpha_{EM}$ is the electromagnetic fine structure
constant. Here the subscripts $L$ and $T$ denote the longitudinal
and transverse polarizations of the virtual photon. The reduced
cross section $\sigma_{r}$ is expressed in terms of the inclusive
proton structure function $F_{2}$ and $F_{L}$ as
\begin{eqnarray}
\sigma_{r}(x,y,Q^{2})=F_{2}(x,Q^{2})-\frac{y^{2}}{1+(1-y)^2}F_{L}(x,Q^{2})
\end{eqnarray}
where $Q^{2}$ is the virtuality of exchanged photon, $\sqrt{s}$
denotes the center of mass energy in $ep$ collisions and
$y=Q^{2}/(sx)$ is the inelasticity variable.\\
The total cross section behavior in some literatures [3,13] is
based on a logarithmic behavior in $x$ which do not violate
unitarity. This behavior has been supplemented by unitarity
correction. In accordance with the Froissart predictions [14],
authors in Ref.[15] have suggested a parameterization  of the
structure function $F_{2}\leq{\ln^{2}(1/x)}$ at large $s$. This
parameterization implies that its growth is limited by the
Froissart bound as Bjorken $x{\rightarrow}0$. This
parameterization will be important in treatments of ultra-high
energy processes. Further investigations of the high-energy limit
of QCD at small $x$ provide valuable information about the
unitarity limits of QCD and parton saturation effects at future
experiments such as an Electron-Ion Collider (EIC) [16], and also
the large Hadron Electron collider (LHeC) [17,18,19] and the
Future Circular Collider program (FCC-eh)[20]. Also some
theoretical analysis at low $x$
have considered the longitudinal structure function $F_{L}(x,Q^{2})$ to describe process [21,22,23].\\
The paper is organized as follows. In sect.2, we give a summary
about the color dipole model. This section reviews the relevant
formulae of the dipole picture. We will study the color dipole
model bounds with respect to the proton structure function $F_{2}$
at low values of $x$ in section 3. It is the purpose of this paper
to predict the behavior of the color dipole bounds at large
$Q^{2}$ values at leading order and next-to-leading order
approximations. I will attempt to preserve analysis of
Altarelli-Martinelli equation and its solution in DGLAP framework
for very small dipoles. In the following the Laplace transform
method of the gluon distribution function with respect to the
transversal and longitudinal structure functions, which both obey
the Froissart boundary conditions, in the LO and NLO
approximations at low values of $x$ are discussed. Explicit
expressions for the Wilson coefficients at the LO are given as
well and at the NLO are given for each $Q^{2}$ values for
simplicity. These parameters at LO and NLO are compared with
bounds in the color dipole model. Section 4 contains the results
and discussions. Numerical results for the extracted these
parameters in the expansion method and Laplace transform method,
together with comparisons with the color dipole bounds are
presented in this section. Also the $W^{2}$ dependence of the
extracted parameters is discussed (where $W^{2}$ is the energy in
the photon-proton center of mass). Conclusions and summary are
summarized on Sec.5. Three Appendices contain results used in the
main text. In Appendix A  the gluon density is discussed with
respect to expansion at distinct points of expansion. Appendix B
explains the steps for obtaining the gluon density by Laplace
transform method. The most cumbersome expressions for the
parameterization of the longitudinal structure function at LO and
NLO
approximations are relegated to Appendix C.\\

\subsection{2. A Short Theoretical Input}

In this section we briefly present the theoretical part of the
color dipole model. For small values of the Bjorken variable  $x$,
the correct degrees of freedom in the high energy $\gamma^{*}p$
scattering ($\gamma^{*}$ emitted by the incident electron) are
given by $q\overline{q}$ colorless dipoles [1]. In the dipole
picture, the $\gamma^{*}p$ interaction process can be described as
follows that first the virtual photon splits into a
$q\overline{q}$ colorless dipole. Then quark and antiquark
interact with proton through radiant gluons. In the leading order,
the generic structure of the $(q\overline{q})p$ is well described
by two-gluon exchange [4]. In deep inelastic scattering (DIS), the
small-$x$ saturation means that the partons in the proton form a
dense system. This system with mutual interaction and
recombination leads to the saturation of the cross section [24].
In the saturation region, the single-gluon exchange changes into
multi-gluon exchange. This process also can be extended to the
next-to-leading order (NLO) corrections as described in
Refs.[25,26]. In Ref.[26] authors discussed the corresponding
corrections in the $\gamma^{*}$ Fock state by adding a new
$q\overline{q}g$ component
to the $q\overline{q}$-state.\\
Within the dipole framework of the $\gamma^{*}p$ scattering
\begin{eqnarray}
\sigma_{L,T}^{\gamma^{*}p}(x,Q^{2})=\int{d^{2}\mathbf{r}}\int
dz\psi^{*}(Q,r,z) \hat{\sigma}(x,r)\psi(Q,r,z).
\end{eqnarray}
Indeed the scattering of the virtual photon on the proton can be
conceived as a virtual photon fluctuation into a quark-antiquark
pair, then the produced quark-antiquark dipole interacts with the
proton via gluon exchanges [7]. The integrands are given by the
squared of the light cone wave functions of the virtual photon and
the scattering amplitudes for the dipole cross section. The first
integration is so-called dipole representation where transverse
momentum $\mathbf{k}_{T}$ is replaced by its Fourier conjugate
variable $\mathbf{r}$. Here $r(\equiv |\mathbf{r}|)$ is the fixed
transverse separation of the quarks in the $q\overline{q}$ pair.
Here the quark (or antiquark) carries a fraction $z$ of the
incoming photon light-cone energy ($0<z<1$). The dipole cross
section $\sigma(x,r)$ is usually assumed to be independent of $z$
and it is a solution of the generalized BFKL equation [27]. Also
it is universal for all flavors and the $x$ dependence  of its
comes from the QCD evolution effects described by the generalized
BFKL equation. The squared wave function of the $q\overline{q}$
Fock states of the  virtual photon is given by the following
equations
\begin{eqnarray}
|\Psi_{T}(z,r)|^{2}&=&\frac{6\alpha_{em}}{4\pi^{2}}\sum_{1}^{n_{f}}e_{f}^{2}\{[z^{2}+(1-z)^{2}]
\epsilon^{2}K_{1}^{2}(\epsilon{r})\nonumber\\
&&+m^{2}_{f}K_{1}^{2}(\epsilon{r})\},\nonumber\\
\mathrm{and}~~~~~~~~~\nonumber\\
|\Psi_{L}(z,r)|^{2}&=&\frac{6\alpha_{em}}{4\pi^{2}}\sum_{1}^{n_{f}}e_{f}^{2}\{4Q^{2}z^{2}(1-z)^{2}K_{0}^{2}(\epsilon{r})\},
\end{eqnarray}
where $n_{f}$ is the number of active quark flavor. In the above
equations $\epsilon^{2}=z(1-z)Q^{2}+m_{f}^{2}$ and $m_{f}$ is the
quark mass. $e_{f}$ is the quark charge and the functions
$K_{0,1}$ are the Bessel-McDonald functions.
 The mass of $q\overline{q}$ dipole is realized by
$M^{2}_{q\overline{q}}=\frac{\overrightarrow{k_{\bot}}^{2}}{z(1-z)}$.
Which the transverse momentum $\overrightarrow{k_{\bot}}$ is
introduced into four momenta of the quark and antiquark.  If the
three momenta
$\overrightarrow{q}=\overrightarrow{k}+\overrightarrow{k}'$ is
defined in the direction of the $z$-axis of a coordinate system,
then the quark and antiquark momenta represented by
\begin{eqnarray}
\overrightarrow{k}=z\overrightarrow{q}+{\overrightarrow{k_{\bot}}},\nonumber\\
\overrightarrow{k}'=(1-z)\overrightarrow{q}-{\overrightarrow{k_{\bot}}}
\end{eqnarray}
where $\overrightarrow{k_{\bot}}.\overrightarrow{q}=0$. With
respect to the center-of-mass energy $W$, the restriction on
masses of the $q\overline{q}$ states is defined by
\begin{eqnarray}
\frac{M^{2}_{q\overline{q}}}{W^{2}}\ll0.1,
\end{eqnarray}
where the Bjorken variable $x{\cong}\frac{Q^{2}}{W^{2}}{\ll}0.1$.
In this approach the photoabsorption cross section  can be
factorized in the following form
\begin{eqnarray}
\sigma_{L,T}^{\gamma^{*}p}(x,Q^{2})&=&\int dz
d^{2}\mathbf{r}_{\bot}
|\Psi_{\gamma}^{L,T}(\mathbf{r}_{\bot},z(1-z),Q^{2})|^{2}\nonumber\\
&&{\times}\widehat{\sigma}_{q\overline{q}}(\mathbf{r}_{\bot},W^{2}),
\end{eqnarray}
where $\sigma_{q\overline{q}}(\mathbf{r}_{\bot},W^{2})$ is the
color-dipole cross-section
\begin{eqnarray}
\widehat{\sigma}_{(q\overline{q})p}(\mathbf{r}_{\bot},W^{2})&=&\int
d^{2}{\overrightarrow{\ell}}_{\bot}
\widetilde{\sigma}_{(q\overline{q})p}({\overrightarrow{\ell}}_{\bot}^{2},W^{2})\nonumber\\
&&{\times}(1-e^{-i{\overrightarrow{\ell}}_{\bot}{\overrightarrow{r}}_{\bot}}),
\end{eqnarray}
which the variable $\mathbf{r}_{\bot}$ determines the transverse
$q\overline{q}$-separation variable and
${\overrightarrow{\ell}}_{\bot}$ stands for the transverse
momentum of the absorbed gluon. In the above integral (i.e.,
Eq.(10)), the first term is associated with the gluon transverse
momentum distribution and the second term is the QCD gauge theory
structure [28,29].\\
On the other hand, the dipole cross section was proposed to have
the following form [3,21]
\begin{eqnarray}
\widehat{\sigma}_{(q\overline{q})p}(x,r)=\sigma_{0}\{1-\exp(-\frac{\pi^{2}r^{2}\alpha_{s}(\mu^{2})xg(x,\mu^{2})}{3\sigma_{0}})
\},
\end{eqnarray}
where $\sigma_{0}$ is a parameter of the model and determined from
a fit to small-$x$ data. This form of the dipole cross section
imposes the unitarity condition at large dipole sizes $r$ as
$\widehat{\sigma}_{(q\overline{q})p}{\leq}\sigma_{0}$. For small
dipole sizes $r$, the dipole cross section is in agreement with
the phenomenon of color transparency resulting from pQCD. The
right-hand side of Eq.(10) is proportional to Eq.(11) in the
small-$r$ region as [28]
\begin{eqnarray}
\alpha_{s}(Q^{2})xg(x,Q^{2})=\frac{3}{4\pi}\int
d{\overrightarrow{\ell}}_{\bot}^{2}{\overrightarrow{\ell}}_{\bot}^{2}
\widetilde{\sigma}_{(q\overline{q})p}({\overrightarrow{\ell}}_{\bot}^{2},W^{2}).
\end{eqnarray}
Plotting the experimental data for $\sigma^{\gamma^{*}p}$ (where
$\sigma^{\gamma^{*}p}=\sigma_{T}^{\gamma^{*}p}+\sigma_{L}^{\gamma^{*}p}=
4\pi^{2}\alpha_{em}F_{2}/Q^{2}$) as a function of the scaling
variable
$\eta(W^{2},Q^{2})=\frac{Q^{2}+m^{2}_{0}}{\Lambda^{2}_{sat}(W^{2})}$
shows a unique behavior as
\begin{equation}
\sigma^{\gamma^{*}p}~{\sim}~\sigma^{(\infty)}
 \begin{cases}
\frac{1}{\eta(W^{2},Q^{2})},~~~~~~~ \mathrm{for} ~\eta\gg1\\
\ln\frac{1}{\eta(W^{2},Q^{2})},~~~~ \mathrm{for} ~\eta\ll1.\\
 \end{cases}
\end{equation}
Here the quantity $\sigma^{(\infty)}$ is independent of the photon
energy, and $\Lambda_{sat}(W^{2})$ is the saturation scale. The
color transparency or saturation of the dipole cross section
depend on that $Q^{2}{\gg}\Lambda_{sat}^{2}(W^{2})$ or
$Q^{2}{\ll}\Lambda_{sat}^{2}(W^{2})$ respectively. Refs.[28] and
[29] show that the saturation scale is defined by
\begin{eqnarray}
\Lambda^{2}_{sat}(W^{2})=\frac{\pi}{\sigma^{(\infty)}}\int
d{\overrightarrow{\ell'}}_{\bot}^{2}{\overrightarrow{\ell'}}_{\bot}^{2}
\widetilde{\sigma}_{(q\overline{q})_{L}^{J=1}}({\overrightarrow{\ell'}}_{\bot}^{2},W^{2}),
\end{eqnarray}
which is fixed spin $J=1$ and $\ell'$ is defined into the gluon
transverse momentum. Also the light-cone variable $z$  reads as
\begin{eqnarray}
{\overrightarrow{\ell'}}_{\bot}^{2}=\frac{{\overrightarrow{\ell}}_{\bot}^{2}}{z(1-z)}.\nonumber
\end{eqnarray}
Therefore, the relationship between gluon distribution and
saturation scale is expressed in the form
\begin{eqnarray}
\alpha_{s}(Q^{2})xg(x,Q^{2})=\frac{1}{8\pi}\sigma^{(\infty)}\Lambda^{2}_{sat}(W^{2}).
\end{eqnarray}
We know that the leading contribution to $F_{2}(x,Q^{2})$ at
sufficiently large $Q^{2}$ in terms of the $J=1$ projection
becomes
\begin{eqnarray}
F_{2}(x,Q^{2})&=&\frac{Q^{2}}{4\pi^{2}\alpha}(\sigma_{\gamma^{*}_{T}p}(W^{2},Q^{2})+\sigma_{\gamma^{*}_{L}p}(W^{2},Q^{2}))\nonumber\\
&&=\frac{R_{e^{+}e^{-}}}{36\pi^{2}}(\int
d{\overrightarrow{\ell'}}_{\bot}^{2}{\overrightarrow{\ell'}}_{\bot}^{2}
\widetilde{\sigma}_{(q\overline{q})_{T}^{J=1}}({\overrightarrow{\ell'}}_{\bot}^{2},W^{2})\nonumber\\
&&+\frac{1}{2}\int
d{\overrightarrow{\ell'}}_{\bot}^{2}{\overrightarrow{\ell'}}_{\bot}^{2}
\widetilde{\sigma}_{(q\overline{q})_{L}^{J=1}}({\overrightarrow{\ell'}}_{\bot}^{2},W^{2})),
\end{eqnarray}
where $R_{e^{+}e^{-}}=3\sum_{f}e_{f}^{2}$. As shown in Ref.[30],
the longitudinal and transverse terms on the right-hand side in
(16) becomes
\begin{eqnarray}
\int
d{\overrightarrow{\ell'}}_{\bot}^{2}{\overrightarrow{\ell'}}_{\bot}^{2}
\widetilde{\sigma}_{(q\overline{q})_{T}^{J=1}}({\overrightarrow{\ell'}}_{\bot}^{2},W^{2})\nonumber\\
=\rho\int
d{\overrightarrow{\ell'}}_{\bot}^{2}{\overrightarrow{\ell'}}_{\bot}^{2}
\widetilde{\sigma}_{(q\overline{q})_{L}^{J=1}}({\overrightarrow{\ell'}}_{\bot}^{2},W^{2}),
\end{eqnarray}
The parameter $\rho$ is associated with the enhanced transverse
size of $q\overline{q}$ fluctuations in the CDM originating from
transverse, $\gamma^{*}_{T}{\rightarrow}q\overline{q}$, and
longitudinal, $\gamma^{*}_{L}{\rightarrow}q\overline{q}$, photons.
Indeed the $\rho$ parameter describes  the ratio of the average
transverse momenta
$\rho=\frac{<\overrightarrow{k}^{2}_{\bot}>_{L}}{<\overrightarrow{k}^{2}_{\bot}>_{T}}$.
It can also be related to the ratio of the effective transverse
sizes of the $(q\overline{q})^{J=1}_{L,T}$ states as
$\frac{<\overrightarrow{r}^{2}_{\bot}>_{L}}{<\overrightarrow{r}^{2}_{\bot}>_{T}}=\frac{1}{\rho}$.
The ratio of the longitudinal to the transversal photoabsorption
cross sections is given by
\begin{eqnarray}
R&=&\frac{\sigma^{\gamma^{*}p}_{L}}{\sigma^{\gamma^{*}p}_{T}}=\frac{1}{2\rho},
\end{eqnarray}
where factor 2 originates from the difference in the photon wave
functions. In terms of the proton structure functions,
$F_{2}(x,Q^{2})$ and $F_{L}(x,Q^{2})$,  the ratio becomes
\begin{eqnarray}
F_{L/2}&\equiv&\frac{F_{L}}{F_{2}}=\frac{1}{1+2\rho}.
\end{eqnarray}
The quantity of $\rho$ in previous analysis [28,29,30,31] is
considered equal to $1$(i.e., $\rho=1$) and for $Q^{2}\gg
\Lambda_{sat}^{2}(W^{2})$ was used the value $\rho=4/3$.  The
deviation from $\rho =1$   quantifies the deviation between the
scatterings  of longitudinally  polarized versus transversally
polarized $q\overline{q}$ fluctuations of the photon. In the
large-$Q^{2}$ limit, the structure function (according to Eq.(14))
takes the form
\begin{eqnarray}
F_{2}(x,Q^{2})&=&\frac{R_{e^{+}e^{-}}}{36\pi^{3}}\sigma^{(\infty)}\Lambda^{2}_{sat}(W^{2})(\rho+\frac{1}{2}).
\end{eqnarray}
The structure function  $F_{2}(x,Q^{2})$ in the small $x$ limit in
the DIS scheme is proportional to the flavor-singlet quark
distribution, $\Sigma(x,Q^{2})$, as
\begin{eqnarray}
F_{2}(x,Q^{2})=\frac{R_{e^{+}e^{-}}}{12}x\Sigma(x,Q^{2}),
\end{eqnarray}
which
$x\Sigma(x,Q^{2})=n_{f}(xq(x,Q^{2})+x\overline{q}(x,Q^{2}))$. Due
to that the sea-quark and gluon distributions have identical
dependence on the kinematic variables, therefore they are assumed
[28,29,30,31] to be proportional to each other with respect to the
$\rho$ parameter as
\begin{eqnarray}
x\Sigma(x,Q^{2})=\frac{8}{3\pi}\alpha_{s}(Q^{2})xg(x,Q^{2})(\rho+\frac{1}{2}).
\end{eqnarray}
Consequently the $\rho$ parameter into the singlet structure
function and gluon distribution function reads as
\begin{eqnarray}
\rho=\frac{3\pi}{8\alpha_{s}(Q^{2})}\frac{F_{2}^{s}(x,Q^{2})}{G(x,Q^{2})}-\frac{1}{2},
\end{eqnarray}
where $F_{2}^{s}(x,Q^{2})=x\Sigma(x,Q^{2})$ and
$G(x,Q^{2})=xg(x,Q^{2})$. Indeed Eq.(23) is valid for very small
dipoles which it is in agreement with the phenomenon of color
transparency resulting from perturbative QCD.  Therefore I will
obtain the relation between gluon distribution function and
dipole cross-section in order to rewrite the ratio of structure functions for the color dipole framework.\\
\subsection{3. Model Description}

In order to describe the ratio of structure functions it is
necessary to make specific methods about the gluon behavior in the
color dipole model upon applying the DGLAP evolution. First, I
give a simple method for the gluon distribution due to the
expansion method at the appropriate points of expansion and
contrast this method with other methods that have appeared
previously. Then I define the gluon distribution and the proton
structure functions according to the  parameterization method.
Then I analyse the consistency between the Altarelli-Martinelli
relation of the gluon distribution and the color dipole model
bounds. In the following I define and discuss the $\rho$ parameter
in the color dipole picture with respect to the  Laplace transform
method at LO and NLO approximations. Finally after the
parameterization of structure functions are specified, I
determined the parameters by rely on the Froissart-bounded
parameterization of the DIS
structure functions.\\

\subsection{3.1. Expansion Method}

 The authors in Refs.[28-31] have used the same correlation between gluon distribution and
derivative of structure function mentioned in Ref.[32] at
sufficiently low values of the Bjorken variable $x\simeq
Q^{2}/W^{2}{\ll}0.1$. The evolution of the structure function with
respect to $\ln{Q^{2}}$ is determined by
\begin{eqnarray}
\frac{\partial F_{2}(x,Q^{2})}{\partial{\ln{Q^{2}}}}\simeq
\frac{\alpha_{s}(Q^{2})}{3\pi}\sum_{q}e_{q}^{2}G(2x,Q^{2}).
\end{eqnarray}
Also another similar relation for the longitudinal structure
function into the gluon density is given  as follows [33]
\begin{eqnarray}
F_{L}(x,Q^{2})\simeq
\frac{2\alpha_{s}(Q^{2})}{5.9\pi}\sum_{q}e_{q}^{2}G(2.5,Q^{2}).
\end{eqnarray}
 Eqs.(24) and (25) actually
indicates that the structure functions are dependent to the gluon
density at low $x$ approximation of the pQCD. That means for
$x\simeq Q^{2}/W^{2}{\ll}1$, the photon-proton interaction is
dominated by the gluon fusion process via
$\gamma^{*}gluon{\rightarrow}q\overline{q}$. In the last years
these methods [32,33] were proposed to isolate the gluon density
by its expansion around $z=\frac{1}{2}$. One method was proposed
in Ref.[34] by the expansion of the gluon density at an arbitrary
$z=a$ (For future discussion please see Appendix A), as
\begin{eqnarray}
\frac{\partial F_{2}(x,Q^{2})}{\partial{\ln{Q^{2}}}}\simeq
\frac{5\alpha_{s}(Q^{2})}{9\pi}\frac{2}{3}G(\frac{x}{1-a}(\frac{3}{2}-a),Q^{2}).
\end{eqnarray}
A similar relationship  for the behavior of the longitudinal
structure function at low $x$  is also described in Ref.[35]
\footnote{These articles [34,35] use the fact that quark densities
can be neglected at low $x$, and the nonsinglet contribution
$F_{2}^{NS}$ can be ignored safely at this limit.}. Therefore the
ratio $\rho$ for a point of expansion $a<1$ can be expressed by
\begin{eqnarray}
\rho=\frac{1}{2}[{F_{2}(x,Q^{2})}/{\frac{\partial
F_{2}(X,Q^{2})}{\partial{\ln{Q^{2}}}}}-1],
\end{eqnarray}
where $X=x\frac{1-a}{\frac{3}{2}-a}$. When the point
$a=\frac{1}{2}$ is used, we get the result expressed in the
literatures [28-31]. Authors in Ref.[35] showed that according to
the expansion method, a general relationship for the longitudinal
structure function into the gluon distribution at low values of
$x$ and at LO approximation can be obtained by the following form
\begin{eqnarray}
F_{L}(x,Q^{2})=
\frac{10\alpha_{s}(Q^{2})}{27\pi}G(\frac{\frac{3}{2}-a}{1-a}x,Q^{2}).\nonumber
\end{eqnarray}
Therefore, we can express Eq.(23) in terms of the longitudinal
structure function as
\begin{eqnarray}
\rho=\frac{1}{2}[{F_{2}(x,Q^{2})}/ F_{L}(X,Q^{2})-1],
\end{eqnarray}
which the longitudinal structure function in connection with the
Froissart bound  at LO and NLO approximations is investigated at
Ref.[21]. In another method, the expansion of the longitudinal
structure function is described into the high order corrections in
Refs.[36,37].\\

\subsection{3.2. Parameterization Method}

Using  a  parameterization suggested by authors in Ref.[15]  on
the proton structure function in a full accordance with the
Froissart predictions [14].  The explicit expression for the
$F_{2}$ parameterization, which obtained from a combined fit of
the H1 and ZEUS collaborations data [38] in a range of the
kinematical variables $x$ and $Q^{2}$( $x<0.01$ and $0.15<
Q^{2}<3000~\mathrm{GeV} ^{2}$), is given by the following form
\begin{eqnarray}
F_{ 2}(x,Q^{2})& =& D(Q^{2})(1-
x)^{n}\sum_{m=0}^{2}A_{m}(Q^{2})L^{m},
\end{eqnarray}
where
\begin{eqnarray}
A_{0}(Q^{2})& =& a_{00} + a_{01}
{\ln}(1+\frac{Q^{2}}{\mu^{2}}),\nonumber\\
 A_{1}(Q^{2})& =& a_{10} + a_{11} {\ln}(1+\frac{Q^{2}}{\mu^{2}}) + a_{12}{\ln}^{2}(1+\frac{Q^{2}}{\mu^{2}})
 ,\nonumber\\
A_{2}(Q^{2})& =& a_{20} + a_{21} {\ln}(1+\frac{Q^{2}}{\mu^{2}}) +
a_{22}{\ln}^{2}(1+\frac{Q^{2}}{\mu^{2}})
 ,\nonumber\\
D(Q^{2})& =& \frac{Q^{2}(Q^{2}+\lambda M^{2})}{(Q^{2}+M^{2})^2},\nonumber\\
L^{m}&=&\ln^{m}(\frac{1}{x}\frac{Q^{2}}{Q^{2}+\mu^{2}}).
\end{eqnarray}
Here $M$ and $\mu^{2}$ are the effective mass  a scale factor
respectively. The additional parameters with their statistical
errors are given in Table I.\\
The point to be considered in Eqs.(24), (25) and (26) is that both
$F_{2}$ and $F_{L}$ are related at small $x$ mainly through the
gluon density. For this purpose, we thoroughly examine the
equations of evolution. According to the  LO DGLAP [40] evolution
equation for 4 massless quarks the formalism introduced in
Refs.[41], the evolution of the proton structure function is given
by
\begin{eqnarray}
\mathcal{F}(x,Q^{2})=x\int_{x}^{1}G(z,Q^{2})K_{qg}(\frac{x}{z})\frac{dz}{z^{2}},
\end{eqnarray}
where $K_{qg}$ is the $\mathrm{gluon}{\rightarrow}\mathrm{quark}$
splitting in leading order of QCD (For more on other quantities,
please see Appendix B). With respect to the Laplace transform
method, the analytical equation for the gluon distribution
$G(x,Q^{2})$ for massless quarks is given by
\begin{eqnarray}
G(x,Q^{2})&=&3\mathcal{FF}(x,Q^{2})-\frac{\partial{\mathcal{FF}(x,Q^{2})}}{\partial{\ln
x
}}\nonumber\\
&&-\int_{x}^{1}\mathcal{FF}(z,Q^{2})(\frac{x}{z})^{3/2}\big{\{}\frac{6}{\sqrt{7}}\sin[\frac{\sqrt{7}}{2}\ln\frac{z}{x}]\nonumber\\
&&+2\cos[\frac{\sqrt{7}}{2}\ln\frac{z}{x}]\frac{dz}{z}\big{\}},
\end{eqnarray}
where
\begin{eqnarray}
\mathcal{FF}(x,Q^{2})=(\frac{\alpha_{s}}{4\pi}\sum_{q}e_{q}^{2})^{-1}\mathcal{F}_{2}(x,Q^{2}),
\end{eqnarray}
and
\begin{eqnarray}
\mathcal{F}_{2}(x,Q^{2}){\equiv}\frac{\partial{F_{2}(x,Q^{2})}}{\partial{\ln
Q^{2}}}-\frac{\alpha_{s}}{4\pi}x\int_{x}^{1}F_{2}(z,Q^{2})K_{qq}(\frac{x}{z})\frac{dz}{z^{2}}.\nonumber\\
\end{eqnarray}
Therefore the ratio $\rho$ is
\begin{eqnarray}
\rho=\frac{27\pi}{20\alpha_{s}(Q^{2})}\frac{F_{2}(x,Q^{2})(\mathrm{i.e.,
Eq.(29)})}{G(x,Q^{2})(\mathrm{i.e., Eq.(32)})}-\frac{1}{2}.
\end{eqnarray}
Indeed the above equation (i.e., Eq.(35)) expressed based on the
Froissart-bounded parameterization of $F_{2}(x,Q^{2})$ while
Eq.(27) is expressed based on the condition of gluon dominant at
low $x$ as gluon carries the $z=a$ fraction from the proton
momentum.\\

\subsection{3.3. Laplace Transform Method at LO approximation}

In the following I developed model due to the Laplace transform
method at LO and NLO approximation. The parameterization of the
structure functions should satisfy the CDM bounds. In Ref.[21] the
longitudinal structure function $F_{L}(x,Q^{2})$ extracted as it
follows the Froissart boundary conditions (For more discussion
please see Appendix C). Now I want to express a new interpretation
for Eq.(23), based on which the ratio $\rho$ will be determined in
terms of $F_{2}(x,Q^{2})$ and $F_{L}(x,Q^{2})$ structure
functions, both of which follow the Froissart boundary condition.
The standard collinear factorization formula for $F_{L}(x,Q^{2})$
at low values of $x$ reads \footnote{Which the non-singlet quark
distribution become negligibly small in comparison with the
singlet distributions.} [54]
\begin{eqnarray}
F_{L}(x,Q^{2})&=&a_{s}(Q^{2})[(c_{L,q}^{(0)}(x)+a_{s}(Q^{2})c_{L,q}^{(1)}(x)+...)\nonumber\\
&&{\otimes}F_{2}(x,Q^{2})+<e^{2}>(c_{L,g}^{(0)}(x)\nonumber\\
&&+a_{s}(Q^{2})c_{L,g}^{(1)}(x)+...){\otimes}G(x,Q^{2})],
\end{eqnarray}
where $<e^{2}>$ is the average squared charge and the
singlet-quark coefficient function is defined by
$c_{L,q}^{(n)}=c_{L,ns}^{(n)}+c_{L,ps}^{(n)}$ which decomposed
into the non-singlet and pure singlet contribution. Some
analytical solutions of the Altarelli-Martinelli [55] equation
have been reported in recent years [56-60] with considerable phenomenological success.\\
Now I use the coordinate transformation in $\upsilon$-space. The
longitudinal structure function reads as
\begin{eqnarray}
\mathcal{\widehat{F}}_{L}(\upsilon,Q^{2})&=&a_{s}(Q^{2})\int_{0}^{\upsilon}[(\widehat{c}_{L,q}^{(n)}(\upsilon-w)+...)
\mathcal{\widehat{F}}_{2}(w,Q^{2})\nonumber\\
&&+<e^{2}>(\widehat{c}_{L,g}^{(n)}(\upsilon-w)+...)\mathcal{\widehat{G}}(w,Q^{2})]dw,\nonumber\\
\end{eqnarray}
where the functions ${\widehat{f}}(\mathrm{i.e.},~
\mathcal{\widehat{F}}, \widehat{c}~\mathrm{and}~
\mathcal{\widehat{G}})$ are defined by
\begin{eqnarray}
\widehat{f}(\upsilon,Q^{2}){\equiv}\widehat{f}(e^{-\upsilon},Q^{2}).\nonumber
\end{eqnarray}
With respect to the Laplace transforms (Please see Appendix B) we
have
\begin{eqnarray}
F_{L}(s,Q^{2})=a_{s}(Q^{2})[k(s) F_{2}(s,Q^{2})+h(s)g(s,Q^{2})],
\end{eqnarray}
where at LO approximation $h(s)=<e^{2}>c_{L,g}^{(0)}(s) $ and
$k(s)=c_{L,q}^{(0)}(s)$. Solving Eq.(38) for $g$, we find that
\begin{eqnarray}
g(s,Q^{2})=\frac{1}{a_{s}(Q^{2})}\frac{F_{L}(s,Q^{2})}{h(s)}-\frac{k(s)}{h(s)}
F_{2}(s,Q^{2}).
\end{eqnarray}
Then we take the inverse laplace transform as the above equation
(i.e., Eq.(39)) can be written as
\begin{eqnarray}
\widehat{G}(\upsilon,Q^{2})&=&\frac{1}{a_{s}(Q^{2})}\mathcal{L}^{-1}
[F_{L}(s,Q^{2}){h(s)}^{-1}]\nonumber\\
&&-\mathcal{L}^{-1} [F_{2}(s,Q^{2})\frac{k(s)}{h(s)}],
\end{eqnarray}
where
$\mathcal{L}^{-1}[g(s,Q^{2});\upsilon]=\widehat{G}(\upsilon,Q^{2})$.
Therefore we find that
\begin{eqnarray}
\widehat{G}(\upsilon,Q^{2})&=&\frac{1}{a_{s}(Q^{2})}\int_{0}^{\upsilon}
\mathcal{\widehat{F}}_{L}(w,Q^{2})\widehat{J}(\upsilon-w)dw\nonumber\\
&&-\int_{0}^{\upsilon}\mathcal{\widehat{F}}_{2}(w,Q^{2})\widehat{L}(\upsilon-w)dw,\nonumber\\
\end{eqnarray}
where $\widehat{J}(\upsilon)$ and $\widehat{L}(\upsilon)$ are new
auxiliary functions, defined by
\begin{eqnarray}
\widehat{J}(\upsilon)&{\equiv}&\mathcal{L}^{-1}[h^{-1}(s);\upsilon],\nonumber\\
\widehat{L}(\upsilon)&{\equiv}&\mathcal{L}^{-1}[{k(s)}{h(s)^{-1}};\upsilon].
\end{eqnarray}
The calculations of $\widehat{J}(\upsilon)$ and
$\widehat{L}(\upsilon)$, using the inverse Laplace transform, are
straightforward and are given in terms of the Dirac delta function
and its derivatives, as we find that
\begin{eqnarray}
\widehat{J}(\upsilon)&=&\frac{3}{4n_{f}}\delta(\upsilon)+\frac{5}{8n_{f}}\delta'(\upsilon)
+\frac{1}{8n_{f}}\delta''(\upsilon),\nonumber\\
\widehat{L}(\upsilon)&=&2C_{F}(\frac{3}{4n_{f}}\delta(\upsilon)+\frac{1}{4n_{f}}\delta'(\upsilon)).
\end{eqnarray}
Using the properties of Dirac delta function, we therefore obtain
an explicit solution for the gluon distribution in terms of the
parameterization of $F_{2}(x,Q^{2})$ [15] and $F_{L}(x,Q^{2})$
[21] by
\begin{eqnarray}
G^{LO}(x,Q^{2})&=&\frac{1}{a_{s}(Q^{2})<e^{2}>}[\frac{1}{8n_{f}}x^{2}\frac{{\partial}^{2}}{{\partial}x^{2}}F_{L}^{LO}(x,Q^{2})\nonumber\\
&&-\frac{5}{8n_{f}}x\frac{{\partial}}{{\partial}x}F_{L}^{LO}(x,Q^{2})+\frac{3}{4n_{f}}F_{L}^{LO}(x,Q^{2})]\nonumber\\
&&-\frac{2C_{F}}{<e^{2}>}[-\frac{1}{4n_{f}}x\frac{{\partial}}{{\partial}x}F_{2}(x,Q^{2})\nonumber\\
&&+\frac{3}{4n_{f}}F_{2}(x,Q^{2})],
\end{eqnarray}
where the parameterization of $F_{2}$ is given by (29). The
explicit expression for the parameterization of $F_{L}$   at the
LO approximation is obtained by the following form [21]
\begin{eqnarray}
F_{L}^{\mathrm{LO}}(x,Q^{2})& =&(1-
x)^{n}\sum_{m=0}^{2}C_{m}(Q^{2})L^{m}.
\end{eqnarray}
The coefficient functions and future discussion about the above
relation can be found in Appendix C. In this formalism both
$F_{2}$ and $F_{L}$ obey the Froissart boundary condition.
Therefore we find the ratio $\rho$ with respect to the Laplace
transform method at LO approximation by the following form
\begin{eqnarray}
\rho=\frac{27\pi}{20\alpha_{s}(Q^{2})}\frac{\mathrm{Eq}.(29)}{\mathrm{Eq}.(44)}-\frac{1}{2}.
\end{eqnarray}

\subsection{3.4. Laplace Transform Method at NLO approximation}

Finally I discuss how the higher-order components of the
coefficient functions may affect these bounds. The longitudinal
structure function within the NLO approximation in
$\upsilon$-space reads as
\begin{eqnarray}
\mathcal{\widehat{F}}_{L}(\upsilon,Q^{2})&=&a_{s}(Q^{2})\int_{0}^{\upsilon}[(\widehat{c}_{L,q}^{(0)}(\upsilon-w)+a_{s}(Q^{2}){\times}\nonumber\\
&&\widehat{c}_{L,q}^{(1)}(\upsilon-w))
\mathcal{\widehat{F}}_{2}(w,Q^{2})+<e^{2}>(\widehat{c}_{L,g}^{(0)}(\upsilon-w)\nonumber\\
&&+a_{s}(Q^{2})\widehat{c}_{L,g}^{(1)}(\upsilon-w))\mathcal{\widehat{G}}(w,Q^{2})]dw.
\end{eqnarray}
Then transform the NLO longitudinal structure function into
$s$-space is
\begin{eqnarray}
F_{L}(s,Q^{2})=a_{s}(Q^{2})[K(s) F_{2}(s,Q^{2})+H(s)g(s,Q^{2})],
\end{eqnarray}
where the coefficient functions at NLO approximation are extended
by $H(s)=<e^{2}>[c_{L,g}^{(0)}(s) +a_{s}(Q^{2})c_{L,g}^{(1)}(s)]$
and $K(s)=c_{L,q}^{(0)}(s)+a_{s}(Q^{2})c_{L,q}^{(1)}(s)$. Now the
inverse Laplace transforms of the above equation (i.e., Eq.(48))
can be performed in the following form
\begin{eqnarray}
\widehat{G}(\upsilon,Q^{2})&=&\frac{1}{a_{s}(Q^{2})}\mathcal{L}^{-1}
[\frac{F_{L}(s,Q^{2})}{<e^{2}>}(c_{L,g}^{(0)}(s)\nonumber\\
&&+a_{s}(Q^{2})c_{L,g}^{(1)}(s))^{-1}]
-\mathcal{L}^{-1}[\frac{F_{2}(s,Q^{2})}{<e^{2}>}\nonumber\\
&&{\times}\frac{c_{L,q}^{(0)}(s)+a_{s}(Q^{2})c_{L,q}^{(1)}(s)}{c_{L,g}^{(0)}(s)
+a_{s}(Q^{2})c_{L,g}^{(1)}(s)}].
\end{eqnarray}
Indeed the gluon distribution at NLO approximation can be
represented as
\begin{eqnarray}
\widehat{G}(\upsilon,Q^{2})&=&\frac{1}{a_{s}(Q^{2})}\int_{0}^{\upsilon}
\mathcal{\widehat{F}}_{L}(w,Q^{2})\widehat{T}(\upsilon-w)dw\nonumber\\
&&-\int_{0}^{\upsilon}\mathcal{\widehat{F}}_{2}(w,Q^{2})\widehat{U}(\upsilon-w)dw,\nonumber\\
\end{eqnarray}
where
\begin{eqnarray}
\widehat{T}(\upsilon)&{\equiv}&\mathcal{L}^{-1}[H^{-1}(s);\upsilon],\nonumber\\
\widehat{U}(\upsilon)&{\equiv}&\mathcal{L}^{-1}[{K(s)}{H(s)^{-1}};\upsilon].
\end{eqnarray}
The inverse Laplace transform of the terms $\widehat{T}(\upsilon)$
and $\widehat{U}(\upsilon)$ at  NLO approximation are
straightforward but they are too lengthy. In the limit case for
these terms, the simplest form can be expressed by the following
form for $Q^{2}=100~\mathrm{GeV}^{2}$ as
\begin{eqnarray}
\widehat{T}(\upsilon)&=&0.2200\delta(\upsilon)+0.1630\delta'(\upsilon)
+0.0320\delta''(\upsilon)\nonumber\\
&&+0.0001e^{-2.0620\upsilon}-0.0036e^{-1.2730\upsilon}\nonumber\\
&&+0.0250e^{-0.5680\upsilon}+0.0189e^{0.0730\upsilon},\nonumber\\
\widehat{U}(\upsilon)&=&0.4980\delta(\upsilon)+0.1580\delta'(\upsilon)+0.0381e^{-2.0620\upsilon}\nonumber\\
&&+0.0620e^{-1.273\upsilon}+0.0040e^{0.0730\upsilon}-0.0210e^{-\upsilon}\nonumber\\
&&-0.0160e^{-0.5680\upsilon}-0.0640e^{-4\upsilon}.
\end{eqnarray}
Therefore an analytical solution for the NLO gluon distribution in
terms of the parameterization of $F_{2}(x,Q^{2})$ [15] and
$F_{L}(x,Q^{2})$ [21] at NLO approximation at
$Q^{2}=100~\mathrm{GeV}^{2}$ is obtained by
\begin{widetext}
\begin{eqnarray}
G^{\mathrm{NLO}}(x,Q^{2})&=&268.361\bigg{[}0.0320~
x^{2}\frac{{\partial}^{2}}{{\partial}x^{2}}F_{L}^{\mathrm{NLO}}(x,Q^{2})
-0.1630~ x\frac{{\partial}}{{\partial}x}F_{L}^{\mathrm{NLO}}(x,Q^{2})+0.2200 F_{L}^{\mathrm{NLO}}(x,Q^{2})\nonumber\\
&&+\int_{x}^{1}\frac{dz}{z}F_{L}^{\mathrm{NLO}}(z,Q^{2})\bigg{\{}0.0001(\frac{x}{z})^{2.0620}
-0.0036(\frac{x}{z})^{1.2720}+0.0250(\frac{x}{z})^{0.5680}+0.0189(\frac{z}{x})^{0.0730}\bigg{\}}\bigg{]}\nonumber\\
&&-\frac{18}{5}\bigg{[}0.4980~ F_{2}(x,Q^{2})-0.1580~
x\frac{{\partial}}{{\partial}x}F_{2}(x,Q^{2})+
\int_{x}^{1}\frac{dz}{z}F_{2}(z,Q^{2})\bigg{\{}0.0381(\frac{x}{z})^{2.0620}
+0.0620(\frac{x}{z})^{1.2720}\nonumber\\
&&-0.0160(\frac{x}{z})^{0.5680}+0.0040(\frac{z}{x})^{0.0730}-0.0210(\frac{x}{z})
-0.0640(\frac{z}{z})^{4}\bigg{\}}\bigg{]},
\end{eqnarray}
\end{widetext}
Therefore, at NLO approximation,  the ratio $\rho$ is obtained as
follows
\begin{eqnarray}
\rho=\frac{27\pi}{20\alpha_{s}(Q^{2})}\frac{\mathrm{Eq}.(29)}{\mathrm{Eq}.(53)}-\frac{1}{2},
\end{eqnarray}
and
\begin{eqnarray}
F_{L/2}^{NLO}=\frac{1}{1+2\rho(\mathrm{Eq}.54)}.\nonumber
\end{eqnarray}
The NLO corrections for the dipole factorization of DIS structure
functions at low $x$ values have been considered in Ref.[61]. In
Ref.[61] the LO approximation for the longitudinal wave-function
of photon is essentially the same as described in the literature
but with an effective vertex. But at NLO approximation, the
colored sector of the virtual photon wave-functions contains both
$q\overline{q}$ and $q\overline{q}g$ components. Expansion of the
structure functions, $F_{2}$ and $F_{L}$, in Fock state in the CDM
are given by
\begin{eqnarray}
F_{2,L}(x,Q^{2})=F_{2,L}^{q\overline{q}}(x,Q^{2})+F_{2,L}^{q\overline{q}g}(x,Q^{2})+...
\end{eqnarray}
The bound $F_{L/2}^{LO}=\frac{1}{3}$ or $\frac{3}{11}$ is valid
only for the first component in the Fock states. Authors in
Ref.[62] showed that at higher Fock states one can be derived the
modified CDM bound for the ratio $F_{L/2}$ as
\begin{eqnarray}
F_{L/2}^{NLO}=F_{L/2}^{LO}\frac{1+\delta\epsilon(x,Q^{2})}{1+\epsilon(x,Q^{2})},
\end{eqnarray}
where
$\epsilon(x,Q^{2})=\frac{F_{2}^{q\overline{q}g}(x,Q^{2})}{F_{2}^{q\overline{q}}(x,Q^{2})}$
and $0{\leq}\delta{\leq}3.7$~. It is assumed that the maximum
$\epsilon$ value is not more than $\%20$.\\
Finally Eq.(54) is my final expression for the ratio $\rho$ within
the NLO approximation for low values of $x$. Indeed I shown that
due to the analysis of the Altarelli-Martinelli equation in DGLAP
approach, the CDM bounds which described for the longitudinal to
transverse ratio at high order correction is valid only for very
small dipoles. The new feature of the model is a parameter
dependent dipole cross section which relying on the
Froissart-bounded parameterization of the structure functions.\\

\subsection{4. Results and Discussions}

QCD evolution effects are taken into account by evolving the gluon
structure function due to the DGLAP evolution and the
Altarelli-Martinelli equation. In the color dipole model the gluon
distribution is modified with respect to the parameterization of
structure functions at LO and NLO approximations. The ratio of the
longitudinal to the transverse photoabsorption cross sections,
$\sigma^{\gamma^{*}p}_{L}/\sigma^{\gamma^{*}p}_{T}$, is extracted.
This result corresponds to the explicit form of the ratio
$F_{2}(x,Q^{2})/G(x,Q^{2})$ by the expansion and Laplace transform
methods. The parameters $\rho$, $R$ and $F_{L/2}$ are obtained
with respect to the the expansion method at LO approximation  and
extended to the NLO approximation due to the Laplace transform
method. The results obtained in the Laplace transform method at LO
and NLO approximations  are dependent on the parameterization of
the structure functions (i.e., $F_{2}$ and $F_{L}$). These
parameterizations [15,21] are valid at low $x$ in a wide range of
the momentum transfer $1<Q^{2}<3000\mathrm{GeV}^{2}$. The active
flavor is selected in these calculations to be equal to $n_{f}=4$
and the QCD parameter $\Lambda$ has been extracted from the
running coupling constant $\alpha_{s}(Q^{2})$ normalized at the
Z-boson mass, as it is chosen [63] to be
$\alpha_{s}(M_{Z}^{2})=0.1166$
from the ZEUS data [64].\\
In Fig.1, the parameters $\rho$, $R$ and $F_{L/2}$ are shown in
terms of the invariant mass $W^{2}(\simeq\frac{Q^{2}}{x})$ in the
interval $10^{3} \mathrm{GeV}^{2}<W^{2}<10^{8} \mathrm{GeV}^{2}$
for three expansion points $a=0.25, 0.50$ and $0.75$. In
traditional literature, the expansion point of the gluon
distribution is chosen by $a=0.50$. Indeed this is the proton
momentum fraction carried by gluon in DIS process. In these
calculations I used the expansion of the gluon distribution at
some arbitrary points and compared with the CDM bounds.\\
Parameters are almost dependent on the invariant mass in the small
expansion points. At high expansion points, the behavior of these
parameters is almost independent of the invariant mass, that
corresponds to the expansion at $a=0.75$. Also a detailed
comparison with the CDM bounds has  been shown  in this figure
(i.e., Fig.1). As can be seen, the values of these parameters are
in good agreement with the CDM bounds in a wide range of the
invariant mass at fixed value of $Q^{2}$. The error bares are in
accordance with the statistical errors of the parameterization of
$F_{2}$
 as presented in Table I.\\
An explicit expression for the gluon density into the proton
structure function at LO approximation by using the Laplace
transform method is derived in Refs.[40]. The result is comparable
to the CTEQ5L [65] and MRST2001LO [66], though there are some
differences with CTEQ5L for large $x$ values. With respect to this
method, a comparison of the parameterization method with expansion
method can be seen for the parameters (i.e., $\rho$, $R$ and
$F_{L/2}$) in Figs.2-4. In the parameterization method, the gluon
obtained by authors in Ref.[41] is used directly. In the expansion
method we use the same $a=0.5$ value used in the literature for
comparison. As can be seen, the invariant mass dependence in high
$Q^{2}$ values is much lower than in low $Q^{2}$ values. According
to the range of errors, it can be seen in these figures that the
results are in the CDM bounds range. Also, the results of
parameterization method are
better than expansion method.\\
Now I proceed with an analysis  of the gluon distribution into the
parameterization of the structure functions as this is of
interests in connection with theoretical investigations of
ultra-high energy processes with cosmic neutrinos. Also this
method is in the context of the Froissart restrictions at low
values of $x$. The longitudinal structure function at low and
mediate $x$ values is written in flavour-singlet quark and gluon
distributions. Therefore in a new method using the Laplace
transform method, the gluon distribution function is expressed in
terms of the structure functions at LO and NLO approximations. The
method relies on the Altarelli-Martinelli equation and on the
Froissart-bounded parameterization of the
 structure functions.\\
In Fig.(5) we present the parameters $\rho$, $R=1/(2\rho)$ and
$F_{L/2}=1/(1+2\rho)$ at LO approximation related to Eq.(46) in
comparison with the
 CDM bounds using the parameterizations of $F_{2}$ and $F_{L}^{LO}$ [15,21].
 As can be seen in this
 figure, one can conclude that the behavior of these parameters
 are almost constant and comparable with the CDM bounds at high $Q^{2}$ values for
 $x\leq 0.01$. Results calculated in LO approximation show that this good comparable
 is only between the bounds and
 results for $100{\leq}Q^{2}{\leq}1000~\mathrm{GeV}^{2}$. These results show that
  the parameters are almost
 $x$ independent for low $x$ values. But they are dependent on $Q^{2}$
 values. However this behavior is consistent with the experimental
 data. However, I need to emphasize that
such results are possible only in a limited kinematics, when
virtuality $Q^{2}$ is very large and significantly exceeds the
saturation scale $Q^{2}_{s}$. These results indicate that the
relationship between gluon PDFs and the dipole cross-section is
consistent with  the CDM bound at high $Q^{2}$ values in order to
rewrite the well-known Altarelli-Martinelli relationship for the
color dipole framework.\\
A very important point that can be seen in all the figures (i.e.,
Figs.1-5) is that the comparability of CDM bounds and results for
large $Q^{2}$ values in these calculations takes place in the
color transparency region where $\eta{\gg}1$. We observe that  for
$Q^{2}=5$ and $10~\mathrm{GeV}^{2}$ we are practically in the
saturation region where $\eta{\ll}1$, which is why the results are
inconsistent with CDM bounds. Figure 6 can be seen to express the
two concepts of saturation and color transparency in the results.
In this figure (i.e., Fig.6) a comparison between the ratio of
structure functions at LO approximation with the H1 data [67] and
the CDM bounds is shown. For
$W^{2}{\gtrsim}10^{4}~\mathrm{GeV}^{2}$, the comparison between
the ratio $F_{L/2}$ and the H1 data with the bounds is good. This
depends on $Q^{2}{\gg}\Lambda^{2}_{sat}(W^{2})$. For
$Q^{2}{\ll}\Lambda^{2}_{sat}(W^{2})$ in the saturation region that
compatibility is not appropriate at all. The results in the color
transparency depend on strong interference between all possible
diagrams. While in the saturation region some diagrams are closed
which causes the photoabsorption cross section to be
${\ln}\Lambda^{2}_{sat}(W^{2})$ dependent, actually turns into the
soft energy dependence. These behaviors are the result of a
general and direct consequence of the color dipole nature of the
interaction of the hadronic fluctuations of the photon with the
color field in the nucleon. Indeed the parameters obtained in the
region $\eta{\gg}1$ can give us good results that can be seen in
all figures. In terms of the nonlinear behavior of the gluon
distribution function, the transition from the region of the
validity of pQCD improves parton model at $\eta{\gg}1$ to the
saturation region of $\eta{\ll}1$ corresponds to a transition from
linear to nonlinear parton distributions. It should be note that
the importance of the saturation regime is when the dipole size
$r$ is significantly larger, $r\sim1/Q_{s}$. In other words, the
formulas mentioned in this article cannot be used in the
saturation regime. For this reason all discussions and final
conclusions apply to the color transparency region.\\
The validity of the results obtained in the LO approximation is
established in the $\eta>1$ region. Now the CDM parameters are
determined and  considered in this region at NLO approximation
with respect to the gluon density in Eq.(53). Results of
calculations of the ratio
$F_{L/2}(W^{2})=\frac{F_{L}(W^{2},Q^{2})}{F_{2}(W^{2},Q^{2})}=\frac{1}{1+2\rho}$
and comparison with the CDM bounds at LO and NLO approximations
are presented in Fig.7. Calculations have been performed at fixed
value $Q^{2}=100~\mathrm{GeV}^{2}$ at low values of $x$, allowing
the invariant mass variable $W^{2}$ to vary in the interval
$10^{3}\mathrm{GeV}^{2}<W^{2}<10^{8}\mathrm{GeV}^{2}$. Figure 7
clearly demonstrates that the ratio extracted is comparable with
the NLO CDM bound. In fact, I have shown that Eq.(22), which
represents the relation between the gluon density and CDM bounds
in the color transparency region, can also be described at NLO
approximation. In conclusion this method defined
consistency between the CDM bounds and gluon PDFs in the color transparency region.\\


\subsection{5. Conclusion}

In this paper the CDM parameters (i.e., $\rho$, $R$ and $F_{L/2}$)
based on one general expression for approximative determination of
the gluon distribution for very small dipoles presented. The gluon
behavior at arbitrary point of expansion of $G(\frac{x}{1-z})$
 is considered. Comparing the
obtained results with the CDM bounds, it can be concluded that the
more suitable points of expansion are in the range $a\geq 0.75$.
Then a method based on the Altarelli-Martinelli equation and on
the Froissart-bounded parameterization of $F_{2}$ and $F_{L}$ with
respect to the Laplace transform method at LO and NLO
approximations when virtuality $Q^{2}$ is  large is proposed. The
parameters obtained on the kinematic region of low and ultra-low
values of $x$ in a large interval of the momentum transfer at LO
approximation. At NLO approximation we focus our attention on
value $Q^{2}=100~\mathrm{GeV}^{2}$. The obtained explicit
expressions for the parameters are entirely determined by the
effective parameters of the $F_{2}$ and $F_{L}$ parameterizations.
The results obtained for the ratio $F_{L/2}$ at NLO approximation
in the color transparency  region are in good agreement with the
NLO CDM bound which  have considered the contribution of the
$q\overline{q}g$ component in the the colored sector of the
virtual photon wave-function.\\
According to the relationships obtained in the color transparency
region at large $Q^{2}$ values, we observe that the results
obtained are comparable to the H1 data and the CDM bounds. These
results indicate that for $Q^{2}{\gg}\Lambda^{2}_{sat}(W^{2})$ the
relationships obtained on the basis of the parameterization of
$F_{2}$ and $F_{L}$ are comparable to the proposed constraints.
The predictions are most reliable for
$20\mathrm{GeV}^{2}{\leq}Q^{2}\leq 1000\mathrm{GeV}^{2}$ at
$x{\leq}0.01$. The behavior of the parameters is independent of
$x$ for large $Q^{2}$ values and are almost dependent on $Q^{2}$
in a wide range of $Q^{2}$ values. They become less reliable, when
$Q^{2}$ decreases to $Q^{2}<20\mathrm{GeV}^{2}$, since in this
case the transition to the saturation region has to be refined by
the nonlinear effects. Indeed in the saturation region the dipole size $r$ is significantly large
 which causes neither DGLAP evolution nor the formulas listed in this paper to be used in the saturated regime.\\


\subsection{ACKNOWLEDGMENTS}

Author is grateful the Razi University for financial support of
this project. The author is especially grateful to N.Nikolaev and
D.Schildknecht for carefully reading
the manuscript and fruitful discussions.\\

\subsection{Appendix A. Details on the Derivation of (26)}

In this Appendix, I provide a brief exposition of the derivation
Eq.(26) in Ref.[34]. The evolution equations for the parton
distributions are defined by
\begin{eqnarray}
\frac{\partial}{\partial{\ln}Q^{2}}f_{i}(x,Q^{2})=P_{ij}(x,Q^{2})
{\otimes}f_{j}(x,Q^{2}),
\end{eqnarray}
where $f_{i}(x,Q^{2})$ stands for the number distributions of
partons in a hadron and $\otimes$ stands for the Mellin
convolution. This equation (i.e., Eq.(57)) represents a system of
$2n_{f}+1$ coupled integro-differential equations. The splitting
functions $P_{ij}(x,Q^{2})$ for $N^{m}LO$ approximation are
defined by
$P^{N^{m}LO}_{ij}(x,Q^{2})=\sum_{k=0}^{m}a_{s}^{k+1}(Q^{2})P_{ij}^{k}(x)$
with $a_{s}(Q^{2})=\alpha_{s}(Q^{2})/4\pi$. The flavor singlet
quark density is defined
$f_{s}=\sum_{i=1}^{n_{f}}[f_{i}+\overline{f}_{i}]$. The LO
evolution equation for $F_{2}$ at low $x$ for four flavors is
defined by
\begin{eqnarray}
\frac{\partial
F_{2}(x,Q^{2})}{\partial{\ln}Q^{2}}=\frac{10\alpha_{s}}{9\pi}
\int_{0}^{1-x}P_{qg}(z) G(\frac{x}{1-z},Q^{2})dz.
\end{eqnarray}
Here the fact is used that at low values of $x$ quark density can
be neglected and the nonsinglet contribution can be ignored. The
authors [34] used the expansion of the gluon distribution at an
arbitrary point $z=a$ as at the limit $x{\rightarrow}0$, the
equation obtained is
\begin{eqnarray}
\frac{\partial F_{2}(x,Q^{2})}{\partial{\ln{Q^{2}}}}\simeq
\frac{5\alpha_{s}(Q^{2})}{9\pi}\frac{2}{3}G(\frac{x}{1-a}(\frac{3}{2}-a),Q^{2}).
\end{eqnarray}
Therefore the gluon distribution can be expressed by
\begin{eqnarray}
G(x,Q^{2})=\frac{9\pi}{5\alpha_{s}(Q^{2})}\frac{3}{2}\frac{\partial
F_{2}(x\frac{1-a}{\frac{3}{2}-a},Q^{2})}{\partial{\ln{Q^{2}}}}.
\end{eqnarray}
The result of comparing them with GRV94(LO) [42] showed that the
better choices have been in the range $0.5{\leq}a{\leq}0.8$ and
with Ryskin $et ~al.$ [43] corresponds to $a=0.75$.\\

\subsection{Appendix B: Details on the Derivation of (32)}

The gluon density used in this analysis obeys the following
Laplace-transform method [44-53], as the coordinate transformation
introduced by $\upsilon{\equiv}{\ln}(1/x)$. Further, Eq.(31)
rewritten by the following form
\begin{eqnarray}
\widehat{\mathcal{F}}(\upsilon,Q^{2})=\int_{0}^{\upsilon}\widehat{G}(w,Q^{2})\widehat{H}(\upsilon-w){dw},
\end{eqnarray}
where
$\widehat{H}(\upsilon){\equiv}e^{-\upsilon}\widehat{K}_{qg}(\upsilon)$
and $\widehat{f}(\upsilon,Q^{2}){\equiv}f(e^{-\upsilon},Q^{2})$ as
$\widehat{K}_{qg}(\upsilon)=1-2e^{-\upsilon}+2e^{-2\upsilon}$. The
Laplace transform of the right-hand of Eq.(61)  is defined by
\begin{eqnarray}
\mathcal{L}\bigg[\int_{0}^{\upsilon}\widehat{G}(w,Q^{2})\widehat{H}(\upsilon-w){dw};s\bigg]=g(s){\times}h(s)
\end{eqnarray}
where
$h(s){\equiv}\mathcal{L}[\widehat{H}(\upsilon);s]=\int_{0}^{\infty}\widehat{H}(\upsilon)e^{-s\upsilon}d\upsilon$
with the condition $\widehat{H}(\upsilon)=0$ for $\upsilon<0$.\\
The gluon distribution function in $\upsilon$-space is obtained in
terms of the inverse transform of a product to the convolution of
the original functions as
\begin{eqnarray}
\widehat{G}(\upsilon,Q^{2})&=&\mathcal{L}^{-1}[f(s,Q^{2}{\times}h^{-1}(s);\upsilon]\nonumber\\
&&=\int_{0}^{\upsilon}\widehat{\mathcal{F}}(w,Q^{2})\widehat{J}(\upsilon-w){dw},
\end{eqnarray}
where
\begin{eqnarray}
\widehat{J}(\upsilon)&{\equiv}&\mathcal{L}^{-1}[h^{-1}(s);\upsilon]\nonumber\\
&&=3\delta(\upsilon)+\delta'(\upsilon)
-e^{-3\upsilon/2}\big{\{}\frac{6}{\sqrt{7}}\sin[\frac{\sqrt{7}}{2}\upsilon]\nonumber\\
&&+2\cos[\frac{\sqrt{7}}{2}\upsilon]\big{\}}.
\end{eqnarray}
Therefore the explicit solution for the gluon distribution in
$\upsilon$-space is defined in terms of the integral
\begin{eqnarray}
\widehat{G}(\upsilon,Q^{2})&=&3\widehat{\mathcal{F}}(\upsilon,Q^{2})+\frac{\partial\widehat{\mathcal{F}}(\upsilon,Q^{2})}
{\partial{\upsilon}}-\int_{0}^{\upsilon}\widehat{\mathcal{F}}(w,Q^{2})\nonumber\\
&&{\times}e^{-3(\upsilon-w)/2}\big{\{}\frac{6}{\sqrt{7}}\sin[\frac{\sqrt{7}}{2}(\upsilon-w)]\nonumber\\
&&+2\cos[\frac{\sqrt{7}}{2}(\upsilon-w)]\big{\}}.
\end{eqnarray}

\subsection{Appendix C: Details about the parameterization of $F_{L}(x,Q^{2})$}
The authors in Ref.[21] obtained  two analytical relations for the
longitudinal structure function at LO and NLO approximations in
terms of the effective parameters of the parameterization of the
proton structure function. The results show that the obtained
method provides reliable longitudinal structure function at HERA
domain and also the structure functions $F_{L}(x,Q^{2})$
manifestly obey the Froissart boundary conditions. The structure
functions, $F_{2}(x,Q^{2})$ and $F_{L}(x,Q^{2})$,  and their
derivatives into $\ln Q^{2}$ are defined with respect to the
singlet and gluon distribution functions $xf_{a}(x,Q^{2})$ as
\begin{eqnarray}
F_{k\{=2,L\}}(x,Q^{2})&=&<e^{2}>\sum_{a=s,g}[B_{k,a}(x){\otimes}xf_{a}(x,Q^{2})]\nonumber\\
\mathrm{and}~~~~~~~~~~~~~~~\nonumber\\
\frac{\partial }{\partial
{\ln}Q^{2}}xf_{a}(x,Q^{2})&=&-\frac{1}{2}\sum_{a,b=s,g}P_{ab}(x){\otimes}xf_{b}(x,Q^{2}).
\end{eqnarray}
The quantities $B_{k,a}(x)$ and $P_{ab}(x)$ are the Wilson
coefficient and splitting functions respectively. The high order
corrections to the coefficient functions can be seen in Ref.[21].
With respect to the Mellin transform method, the leading order
longitudinal structure function is obtained at low $x$ by the
following form
\begin{eqnarray}
F_{L}^{\mathrm{LO}}(x,Q^{2})& =&(1-
x)^{n}\sum_{m=0}^{2}C_{m}(Q^{2})L^{m},
\end{eqnarray}
where
\begin{eqnarray}
C_{2}&=&\widehat{A}_{2}+\frac{8}{3}a_{s}(Q^{2})DA_{2}\nonumber\\
C_{1}&=&\widehat{A}_{1}+\frac{1}{2}\widehat{A}_{2}+\frac{8}{3}a_{s}(Q^{2})D[A_{1}+(4\zeta_{2}-\frac{7}{2})A_{2}]\nonumber\\
C_{0}&=&\widehat{A}_{0}+\frac{1}{4}\widehat{A}_{2}-\frac{7}{8}\widehat{A}_{2}+\frac{8}{3}a_{s}(Q^{2})D[A_{0}
+(2\zeta_{2}-\frac{7}{4})A_{1}\nonumber\\
&&+(\zeta_{2}-4\zeta_{3}-\frac{17}{8})A_{2}],
\end{eqnarray}
and
\begin{eqnarray}
\widehat{A}_{2}&=&\widetilde{A}_{2}\nonumber\\
\widehat{A}_{1}&=&\widetilde{A}_{1}+2DA_{2}\frac{\mu^{2}}{\mu^{2}+Q^{2}}\nonumber\\
\widehat{A}_{0}&=&\widetilde{A}_{0}+DA_{1}\frac{\mu^{2}}{\mu^{2}+Q^{2}}\nonumber\\
\widetilde{A}_{i}&=&\widetilde{D}A_{i}+D\overline{A}_{i}\frac{Q^{2}}{Q^{2}+\mu^{2}}\nonumber\\
\widetilde{D}&=&\frac{M^{2}Q^{2}[(2-\lambda)Q^{2}+\lambda
M^{2}]}{[Q^{2}+M^{2}]^{3}}\nonumber\\
\overline{A}_{m}&=&a_{m1}+2a_{m2}L_{2},~~a_{02}=0.
\end{eqnarray}
The standard representation for QCD running coupling constant in
the LO and NLO approximations have been described
\begin{eqnarray}
a_{s}^{\mathrm{LO}}(Q^{2})&=&\frac{1}{\beta_{0}\ln(Q^{2}/\Lambda^{2})},\\
a_{s}^{\mathrm{NLO}}(Q^{2})&=&\frac{1}{\beta_{0}\ln(Q^{2}/\Lambda^{2})}-\frac{\beta_{1}{\ln}\ln(Q^{2}/\Lambda^{2})}{\beta_{0}[\beta_{0}\ln(Q^{2}/\Lambda^{2})]^{2}},\nonumber
\end{eqnarray}
which the QCD parameter $\Lambda$ at LO and NLO approximations has
been extracted with $\Lambda^{LO}(n_{f}=4)=136.8~\mathrm{MeV}$ and
$\Lambda^{NLO}(n_{f}=4)=284.0~\mathrm{MeV}$ respectively. The QCD
$\beta$-functions are
\begin{eqnarray}
\beta_{0}=\frac{1}{3}(11C_{A}-2n_{f}),~\beta_{1}=\frac{1}{3}(34C_{A}^{2}-2n_{f}
(5C_{A}+3C_{F}))\nonumber
\end{eqnarray}
where $C_{F}$ and $C_{A}$ are the Casimir operators in the
$SU(N_{c})$ color group.\\
The NLO longitudinal structure function at small $x$ is defined by
the following form
\begin{eqnarray}
F_{L}^{\mathrm{NLO}}(x,Q^{2})&=&\frac{1}{[1+\frac{1}{3}a_{s}(Q^{2})L_{C}
(\widehat{\delta}^{(1)}_{sg}-\widehat{R}^{(1)}_{L,g})]}\bigg{\{}\nonumber\\
&&[1-a_{s}(Q^{2})
(\overline{\delta}^{(1)}_{sg}-\overline{R}^{(1)}_{L,g})]F_{L}^{\mathrm{LO}}(x,Q^{2})\nonumber\\
&&-a^{2}_{s}(Q^{2})[\frac{1}{3}\widehat{B}^{(1)}_{L,s}L_{A}+\overline{B}^{(1)}_{L,s}]F_{2}(x,Q^{2})
\bigg{\}}\nonumber\\
\end{eqnarray}
where the coefficient functions read as
\begin{eqnarray}
\widehat{B}^{(1)}_{L,s}&=&8C_{F}[\frac{25}{9}n_{f}-\frac{449}{72}C_{F}+(2C_{F}-C_{A})\nonumber\\
&&(\zeta_{3}+2\zeta_{2}-\frac{59}{72})]\nonumber\\
\overline{B}^{(1)}_{L,s}&=&\frac{20}{3}C_{F}(3C_{A}-2n_{f})\nonumber\\
\widehat{\delta}^{(1)}_{sg}&=&\frac{26}{3}C_{A}\nonumber\\
\overline{\delta}^{(1)}_{sg}&=&3C_{F}-\frac{347}{18}C_{A}\nonumber\\
\widehat{R}^{(1)}_{L,g}&=&-\frac{4}{3}C_{A}\nonumber\\
\overline{R}^{(1)}_{L,g}&=&-5C_{F}-\frac{4}{9}C_{A}\nonumber\\
L_{A}&=&L+\frac{A_{1}}{2A_{2}}\nonumber\\
L_{C}&=&L+\frac{C_{1}}{2C_{2}}\nonumber\\
L&=&\ln(1/x)+L_{1}\nonumber\\
L_{1}&=&{\ln}\frac{Q^{2}}{Q^{2}+\mu^{2}}.
\end{eqnarray}
\begin{table}[h]
\caption{ The effective Parameters at low $x$ for
$0.15~\mathrm{GeV}^{2}<Q^{2}<3000~\mathrm{GeV}^{2}$ provided by
the following values. The fixed  parameters are defined by the
Block-Halzen fit [39] to the real photon-proton cross section as
$M^{2}=0.753 \pm 0.068~ \mathrm{GeV}^{2}$ and $\mu^2 = 2.82 \pm
0.290~ \mathrm{GeV}^{2}$.}
\begin{tabular} {cccc}
\toprule \\  \multicolumn{2}{c}{parameters \quad \quad \quad ~~~~~~~~~~~~~~~~value}    \\ &&&\\ \hline \\ &&&\\
$a_{00}$& \quad  $2.550\times 10^{-1}~\pm 1.60\times10^{-2}$ \\

$a_{01}$& \quad  $1.475\times 10^{-1}~\pm 3.025\times10^{-2}$\\&&&\\

  $a_{10} $  &   \quad  $8.205\times 10^{-4}~~  \pm  4.62\times10^{-4} $  \\

  $a_{11} $  &   \quad   $-5.148\times 10^{-2}\pm 8.19\times10^{-3}$  \\

  $a_{12}$   &    \quad  $-4.725\times 10^{-3}\pm 1.01\times10^{-3}$   \\  &&&\\

 $a_{20}$   &   \quad   $2.217\times 10^{-3}\pm 1.42\times10^{-4} $ \\

 $a_{21}$   &   \quad   $1.244\times 10^{-2}\pm 8.56\times10^{-4}$  \\

 $a_{22}$    &    \quad  $5.958\times 10^{-4}\pm 2.32\times10^{-4} $ \\ &&& \\

$n$& \quad  $11.49\pm 0.99$ & &\\

$\lambda$& \quad  $2.430~\pm 0.153$ & &\\

$\chi^{2}(\mathrm{goodness~ of~ fit})$ &  \quad  $0.95$ & &\\

\hline

\end{tabular}
\end{table}
\newpage{
\section{References}
1. N.N.Nikolaev and B.G.Zakharov, Z.Phys.C{\bf49}, 607(1991);
N.N.Nikolaev and B.G.Zakharov, Z.Phys.C{\bf53}, 331(1992);
A.H.Mueller, Nucl.Phys.B{\bf415}, 373(1994); Nucl.Phys.B{\bf437},
107(1995); K.Golec-Biernat, Acta
Phys.Pol.B{\bf35}, 3103(2004).\\
2. R.J.Glauber, Phys.Rev.{\bf99}, 1515(1955); A.H.Mueller,
Nucl.Phys.B{\bf335}, 115(1990).\\
3. K.Golec-Biernat and M.W$\mathrm{\ddot{u}}$sthoff,
Phys.Rev.D{\bf59}, 014017(1998); K.Golec-Biernat and
M.W$\mathrm{\ddot{u}}$sthoff, Phys.Rev.D{\bf60}, 114023(1999).\\
4. K.Golec-Biernat, Acta Phys.Pol.B{\bf33}, 2771(2002);
J.Phys.G{\bf28}, 1057(2002); H.Kowalski and D.Teaney,
Phys.Rev.D{\bf68}, 114005(2003); H.Kowalski, L.Motyka and
G.Watt, Phys.Rev.D{\bf74}, 074016(2006).\\
5. C.Ewerz, A.von Manteuffel and O.Nachtmann, JHEP{\bf03},
102(2010); D.Britzger et al., Phys.Rev.D{\bf100}, 114007 (2019); C.Ewerz, A. von Manteuffel
 and O.Nachtmann, Phys.Rev.D{\bf77}, 074022(2008).\\
6. H.Kowalski and D.Teaney, Phys.Rev.D{\bf68}, 114005(2003).\\
7. G.Watt and H.Kowalski,  Phys.Rev.D{\bf78}, 014016(2008);
 A.H.Rezaeian et al., Phys.Rev.D{87}, 034002(2013).\\
8. L.McLerran, arXiv:0804.1736 [hep-ph] (2008);
 Acta.Phys.Polon.B{\bf45}, 2307 (2014).\\
9. F. Gelis et al.,
 Annu. Rev. Nucl. Part. Sci. {\bf60}, 463 (2010); G.M.Peccini et al., Phys. Rev. D{\bf101}, 074042 (2020);
  M.Genovese, N.N.Nikolaev and B.G.Zakharov, J.Exp.Theor.Phys.{\bf81}, 633(1995); J.Exp.Theor.Phys.{\bf81}, 625(1995);
   Amir H.Rezaeian and I.Schmidt, Phys.Rev. D{\bf88}, 074016 (2013).\\
10. I.Balitsky, Nucl.Phys.B{\bf463}, 99(1996); Phys.Rev.D{\bf75},
014001(2007); Y.V.Kovchegov, Phys.Rev.D{\bf60}, 034008(1999);
Phys.Rev.D{\bf61}, 074018(2000).\\
11. E.Iancu,K.Itakura and S.Munier, Phys.Lett.B{\bf590},
199(2004).\\
12. A.M.Stasto, K.Golec-Biernat, J.Kwiecinski,
Phys.Rev.Lett.{\bf86}, 596(2001); F.Gelis et al.,
Phys.Lett.B{\bf647}, 376(2007); J.Kwiecinski and A.M.Stasto, Phys.Rev.D{\bf66}, 014013(2002).\\
13. D.Schildknecht and H.Spiesberger, arXiv [hep-ph]: 9707447
(1997); W.Buchmuller
and D.Haidt, arXiv [hep-ph]: 9605428  (1996) .\\
14. M.Froissart, Phys.Rev.{\bf123}, 1053(1961).\\
15. M. M. Block, L. Durand and P. Ha, Phys. Rev.D{\bf 89}, no. 9,
094027 (2014); M.M.Block et al., Phys. Rev.D{\bf 88}, no. 1,
014006 (2013).\\
16. D.Boer et al., arXiv: [nucl-th]1108.1713.\\
17. M.Klein,  arXiv [hep-ph]:1802.04317; M.Klein,
Ann.Phys.{\bf528}, 138(2016); N.Armesto et al.,
Phys.Rev.D{\bf100}, 074022(2019).\\
18. J.Abelleira Fernandez et al., [LHeC Study Group
Collaboration], J.Phys.G{\bf39}, 075001(2012).\\
19. P.Agostini et al. [LHeC Collaboration and FCC-he Study Group
], CERN-ACC-Note-2020-0002, arXiv:2007.14491 [hep-ex] (2020).\\
20. A. Abada et al., [FCC Study Group Collaboration], Eur.Phys.J.C{\bf 79}, 474(2019).\\
21. L.P.Kaptari et al., Phys.Rev.D{\bf99}, 096019 (2019).\\
22. B.Rezaei and G.R.Boroun, Phys.Rev.C{\bf101}, 045202 (2020).\\
23. G.R.Boroun and B.Rezaei, Nucl.Phys.A{\bf990}, 244(2019).\\
24 K.Golec-Biernat and S.Sapeta, JHEP {\bf03}, 102 (2018); J.Bartels, K.Golec-Biernat and H.Kowalski, Phys.Rev.D{\bf66}, 014001(2002).\\
25. G.Beuf, Phys.Rev.D{\bf94}, 054016(2016); Phys.Rev.D{\bf96},
074033 (2017); G.Beuf et al., arXiv:2008.05233[hep-ph] (2020).\\
26. B.Duclou$\mathrm{\acute{e}}$ et al., Phys.Rev.D{\bf96},
094017(2017); J.Bartels et al., Phys.Rev.D{\bf81}, 054017(2010).\\
27. N.N.Nikolaev and B.G.Zakharov, Phys.Lett.B{\bf333}, 250(1994);
L.Frankfurt, A.Radyushkin and M.Strikman, Phys.Rev.D{\bf55},
98(1997); M. Genovese, N.N. Nikolaev and  B.G. Zakharov, JETP
{\bf81}, 633(1995); I.P. Ivanov and  N.N. Nikolaev, Phys. Rev. D
{\bf65}, 054004 (2002).\\
28. M.Kuroda and D.Schildknecht, Phys.Lett. B{\bf618}, 84(2005);
M.Kuroda and D.Schildknecht, Acta Phys.Polon. B{\bf37}, 835(2006);
 M.Kuroda and D.Schildknecht, Phys.Lett. B{\bf670}, 129(2008); M.Kuroda and D.Schildknecht, Phys.Rev. D{\bf96}, 094013(2017).\\
29. D.Schildknecht and M.Tentyukov, arXiv[hep-ph]:0203028;
M.Kuroda and D.Schildknecht, Phys.Rev. D{\bf85}, 094001(2012);
 D.Schildknecht, Mod.Phys.Lett.A{\bf 29}, 1430028(2014);
 M.Kuroda and D.Schildknecht, Int. J. Mod. Phys. A{\bf31}, 1650157 (2016).\\
30.  M.Kuroda and D.Schildknecht, Phys.Rev.D{\bf31}, 094001(2014).\\
31. D.Schildknecht, B.Surrow and M.Tentynkov,  Phys.Lett.B{499},
 116(2001); G.Cvetic , D.Schildknecht, B.Surrow and M.Tentynkov,
 Eur.Phys.J.C{\bf20}, 77(2001); D.Schildknecht, B.Surrow and M.Tentynkov,  Mod.Phys.Lett.A{16},
 1829(2001); M.Kuroda and D.Schildknecht, Eur.Phys.J.C{37},
 205(2004).\\
32. K.Prytz, Phys.lett.B{\bf311}, 286(1993).\\
33. A.M.Cooper-Sarkar et al., Z.Phys.C{\bf39}, 281(1988);
A.M.Cooper-Sarkar et al., Acta Phys.Pol.B{\bf34}, 2911(2003).\\
34. M.B.Gay Ducati and Victor P.B.Goncalves, Phys.Lett.B{\bf390}, 401(1997).\\
35. G.R.Boroun and B.Rezaei, Eur.Phys.J.C{72}, 2221(2012).\\
36. B.Rezaei and G.R.Boroun, Eur.Phys.J.A{\bf56}, 262 (2020).\\
37. G.R.Boroun and B.Rezaei, Phys.Lett.B {\bf816}, 136274 (2021).\\
38. F.D.Aaron et al., [H1 and ZEUS Collaborations], JHEP{\bf1001},
109(2010).\\
39. M.M.Block and F.Halzen, Phys.Rev.Lett.{\bf107}, 212002 (2011);
Phys.Rev.D{\bf70}, 091901 (2004).\\
40. Yu.L.Dokshitzer, Sov.Phys.JETP {\textbf{46}}, 641(1977);
G.Altarelli and G.Parisi, Nucl.Phys.B \textbf{126}, 298(1977);
V.N.Gribov and L.N.Lipatov, Sov.J.Nucl.Phys. \textbf{15},
438(1972).\\
41. Martin M.Block, Eur.Phys.J.C{\bf65}, 1 (2010); M. M. Block, L.
Durand and Douglas W.McKay, Phys. Rev.D{\bf 79},  0140131 (2009);
Phys. Rev.D{\bf 77},  094003 (2008).\\
42. M.Gluk, E.Reya and A.Vogt, Z.Phys.C{\bf67}, 433(1995).\\
43. M.G.Ryskin, Yu. M.Shabelski and A.G.Shuvaev, Z.Phys.C{\bf73}, 111(1996).\\
44. M.M.Block, L.Durand and D.W.McKay, Phys.Rev.D{\bf79}, 014031(2009);  M.M.Block, Eur.Phys.J.C{\bf65}, 1(2010).\\
45. F.Taghavi-Shahri, A.Mirjalili and M.M.Yazdanpanah,
Eur.Phys.J.C{\bf71}, 1590(2011).\\
46. S.M.Moosavi Nejad et al., Phys.Rev.C{\bf94}, 045201(2016).\\
47. H.Khanpour, A.Mirjalili and S.Atashbar Tehrani,
Phys.Rev.C{\bf95}, 035201(2017).\\
48. G.R. Boroun, S. Zarrin and F. Teimoury,
Eur.Phys.J.Plus{\bf130}, 214(2015).\\
49. G.R. Boroun and S. Zarrin, Phys.Atom.Nucl.{\bf78},
1034(2015).\\
50. G.R. Boroun, S. Zarrin and S. Dadfar, Nucl.Phys.A{\bf953},
21(2016).\\
51. G.R. Boroun, S. Zarrin and S. Dadfar, Phys.Atom.Nucl.{\bf79},
236(2016).\\
52. F. Teimoury Azadbakht, G.R. Boroun and B. Rezaei,
Int.J.Mod.Phys.E{\bf27}, 1850071(2018).\\
53. S.Dadfar and S.Zarrin, Eur.Phys.J.C{\bf80},
319(2020).\\
54. S.Moch, J.A.M.Vermaseren and A.Vogt, Phys.Lett.B{\bf606},
123(2005); D.I.Kazakov et al., Phys.Rev.Lett.{\bf65}, 1535(1990).\\
55. G.Altarelli and G.Martinelli, Phys.Lett.B{\bf76}, 89(1978).\\
56. N.Baruah, M.K.Das and J.K.Sarma, Few-Body Syst.{\bf55},
1061(2014); N.Baruah, N.M.Nath and J.K.Sarma,
Int.J.Theor.Phys.{\bf52}, 2464(2013).\\
57. G.R.Boroun, Phys.Rev.C{\bf97}, 015206(2018).\\
58. G.R.Boroun, Int.J.Mod.Phys.E{\bf18}, 131(2009).\\
59. G.R.Boroun, B.Rezaei and J.K.Sarma, Int.J.Mod.Phys.A{\bf29},
1450189(2014).\\
60. G.R.Boroun,
Eur.Phys.J.Plus{\bf129}, 19(2014).\\
61. G.Beuf,  Phys. Rev. D {\bf85}, 034039(2012).\\
62. M.Niedziela and M.Praszalowicz, Acta Physica Polonica
B{\bf46}, 2018(2015).\\
63. K.G.Chetyrkin, B.A.Kniehl and M.Steinhauser,
Phys.Rev.Lett.{\bf79}, 2184(1997); Nucl.Phys.B{\bf510},
61(1998).\\
64. S.Chekanov et al.[ZEUS Collaboration],
Eur.Phys.J.C{\bf21},443(2001).\\
65. H.L.Lai et.al. [CTEQ Collaboration], Eur.Phys.J.C{\bf12},
375(2000).\\
66. A.D.Martin, R.G.Roberts, W.J.Stirling and R.S.Thorne,
MRST2001, Eur.Phys.J.C{\bf23}, 73(2002).\\
67. V.Andreev et al. [H1 Collaboration], Eur.Phys.J.C{\bf74},
2814(2014).\\
}
\begin{figure}
\includegraphics[width=0.55\textwidth]{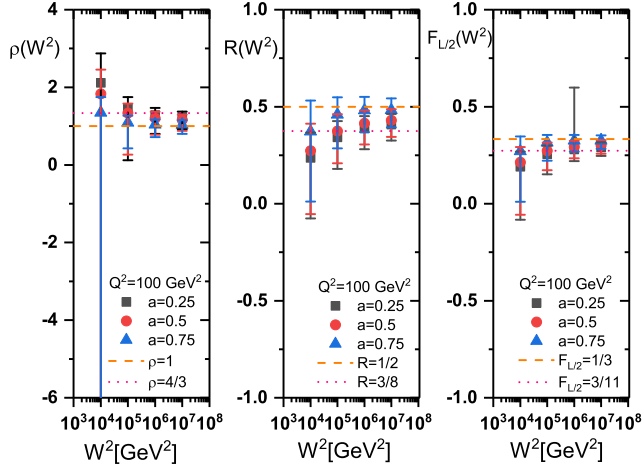}
\caption{Results of the parameters $\rho, R$ and $F_{L/2}$
obtained from the three expansion points $a=0.25, 50$ and $0.75$
 respectively. The parameters
compared with the CDM bounds $\rho=1$ and $4/3$, $R=1/2$ and $3/8$
and $F_{L/2}=1/3$ and $3/11$ respectively. The results are
presented  at fixed $Q^{2}$ value  in the interval
$10^{3}\mathrm{GeV}^{2 }< W^{2} < 10^{8}\mathrm{GeV}^{2}$ at low
values of $x$. The error bars are correspondent to the
uncertainties of the $F_{2}$ parameterization (i.e., Table I).
}\label{Fig1}
\end{figure}
\begin{figure}
\includegraphics[width=0.55\textwidth]{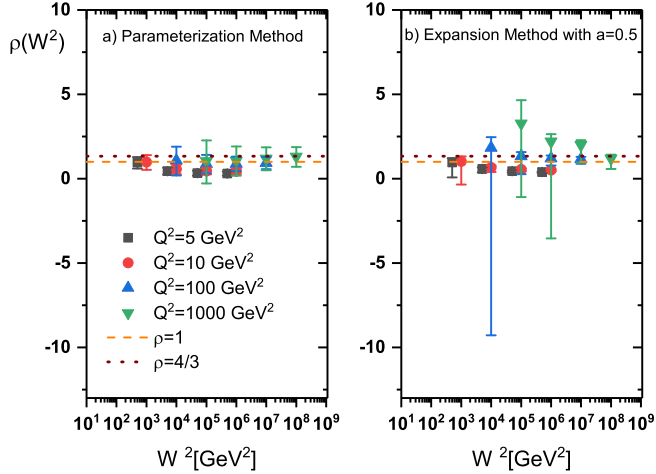}
\caption{Comparison between the results obtained in
parameterization method and expansion method at $a=0.5$ for the
$\rho$ parameter. The obtained values compared with the CDM bounds
$\rho=1$ and $4/3$. The results are presented  at four fixed
values of $Q^{2}$ ($Q^{2}=5,~ 10,~ 100~ \mathrm{and}~
1000~\mathrm{GeV}^{2}$) in the interval $10^{2}\mathrm{GeV}^{2 }<
W^{2} < 10^{8}\mathrm{GeV}^{2}$. The error bars are correspondent
to the uncertainties of the $F_{2}$ parameterization (i.e., Table
I).}\label{Fig2}
\end{figure}
\begin{figure}
\includegraphics[width=0.55\textwidth]{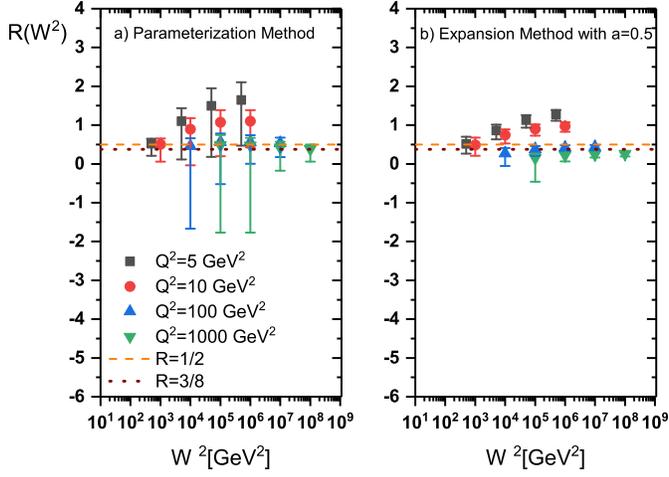}
\caption{The same as Fig.2 for the ratio $R$ which compared with
the CDM bounds $R=1/2$ and $3/8$. }\label{Fig3}
\end{figure}
\begin{figure}
\includegraphics[width=0.55\textwidth]{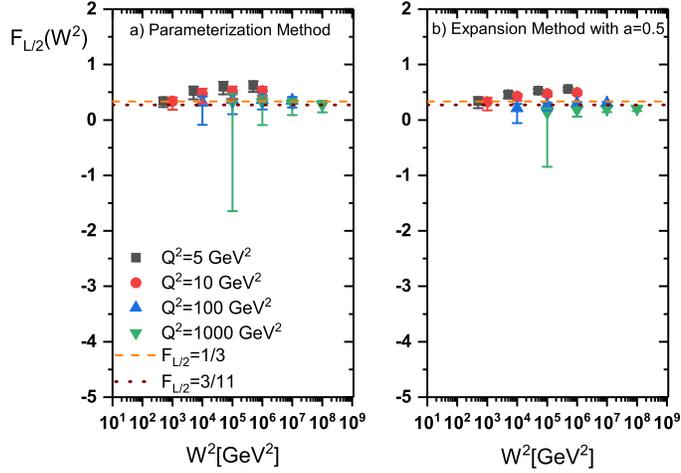}
\caption{The same as Fig.2 for the ratio $F_{L/2}$ which compared
with the CDM bounds $F_{L/2}=1/3$ and $3/11$.}\label{Fig4}
\end{figure}
\begin{figure}
\includegraphics[width=0.55\textwidth]{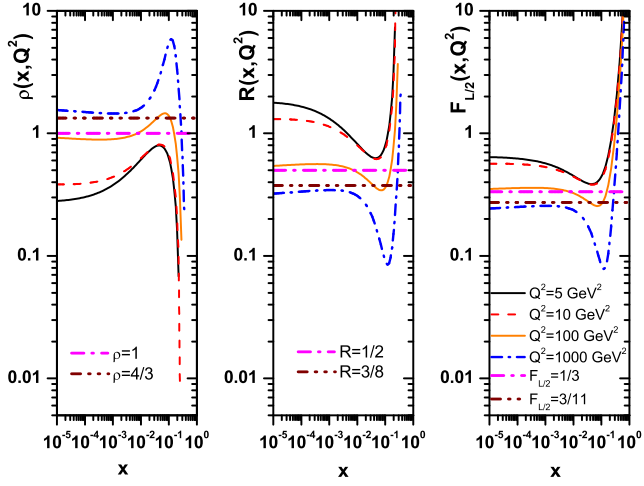}
\caption{Parameters $\rho$, $R$ and $F_{L/2}$ at LO approximation
as a function of  $x$  for $Q^{2}=5,~10, ~100$ and
$1000~\mathrm{GeV}^{2}$ compared with the CDM bounds.}\label{Fig5}
\end{figure}
\begin{figure}
\includegraphics[width=0.55\textwidth]{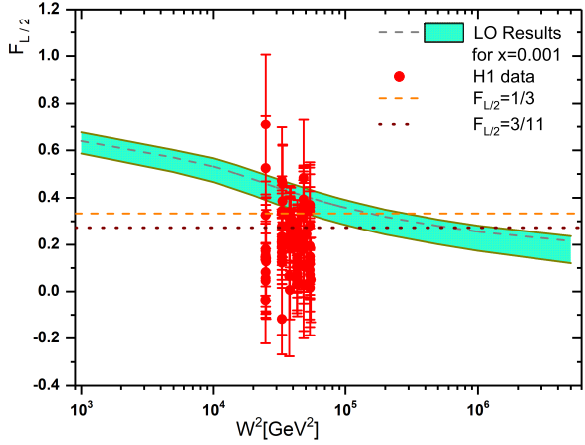}
\caption{The ratio of the longitudinal to transversal structure
functions calculated within the LO approximation at fixed value of
the Bjorken variable $x=0.001$. Experimental data are from the
H1-Collaboration as accompanied with total errors, Ref.[65]
corresponding to the chosen kinematics lies in the interval
$1.5~\mathrm{GeV}^{2}{\leq}Q^{2}{\leq}150~\mathrm{GeV}^{2}$. The
obtained values compared with the CDM bounds $F_{L/2}=1/3$ and
$3/11$. The error bars are correspondent to the uncertainties of
the $F_{2}$ and $F_{L}^{LO}$ parameterization (i.e., Table I and
Appendix C).}\label{Fig6}
\end{figure}
\begin{figure}
\includegraphics[width=0.5\textwidth]{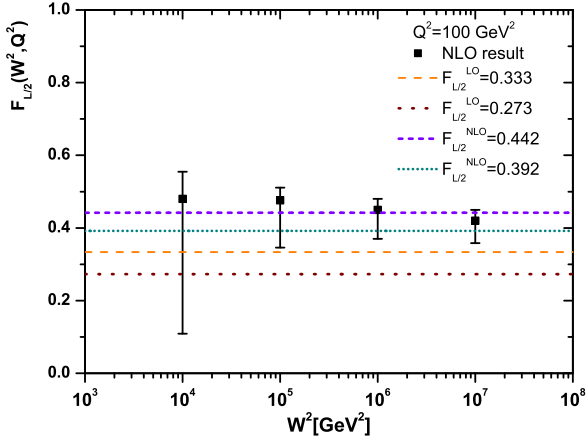}
\caption{Ratios $F_{L/2}$ plotted as function of the invariant
mass $W^{2}$. Straight lines correspond to the CDM bounds at LO
and NLO approximations. The obtained values at NLO approximation
compared with the  NLO CDM bounds $F_{L/2}=0.392$ and $0.442$
[67]. The error bars are correspondent to the uncertainties of the
$F_{2}$ and $F_{L}^{NLO}$ parameterization (i.e., Table I and
Appendix C).}\label{Fig7}
\end{figure}
\end{document}